# Dynamic Variation in Protein-Small Molecule Interaction Observed by Double-Nanohole Optical Trapping

*Ahmed Al Balushi, Reuven Gordon\**

*Department of Electrical Engineering, University of Victoria, Victoria, British Columbia V8W 3P6, Canada*

ABSTRACT: The interaction of proteins with small molecules is fundamental to their function in living organisms and it is widely studied in drug development. Here we compare optical trapping dynamics of streptavidin and biotinylated streptavidin using a double nanohole optical trap in a metal film. Consistent and clearly distinct behavior is seen between the protein with and without the small molecule binding. The real-time dynamics at the single protein level are accessible with this technique, which also has advantages of not requiring tethering to a surface or the need for exogenous markers.

Protein-small molecule interactions (PSMIs) play an important role in biological functions**.** PSMIs are also of primary interest for the development of drugs, for example through inhibition of protein interactions.[1] While many works have studied PSMIs, only few approaches exist that do not require tethering to a surface or labelling. Tethering to a surface has the disadvantages of using a binding site, restricting the protein motion with an anchor and introducing steric hindrance from the surface proximity. Exogenous labels, such as fluorescent tags, present similar challenges, including using up a binding site and altering the natural state of the molecules of interest, but also add the complexity of using a label. So far, calorimetry[2] and interferometry[3] have been used as label-free, free-solution techniques. Calorimetry has high concentration detection limits and is restricted to systems with an appreciable reaction enthalpy. Interferometry makes use of refractive index changes from PSMIs and can detect PSMIs in the micromolar range.

Ideally, we would like to introduce new label-free, free-solution methods that work at the single molecule level, where it is possible to observe real time dynamics that are not obscured by an ensemble. There are many advantages of working at the single molecule level.[4] For example, with access to these dynamics and without the need for synchronization, we can compare more directly to molecular dynamics calculations (for example, Ref. 5). Working at the single molecule level represents the ultimate practical sensitivity limit. It also offers opportunities for distinguishing components of heterogeneous systems, such as cell lysates.

Here we consider the use of optical trapping to observe, in real time, the dynamics of a single streptavidin molecule, comparing the cases with and without exposure to biotin. Our optical trapping approach is similar to that reported in previous works,[6-10] as shown schematically in Figure 1(a). Briefly, a double nanohole aperture is milled using a focused ion beam in a 100 nm Au film adhered to a glass slide with a 5 nm Ti layer (Fig. 1(c)). The gold film forms the top of a microwell in an inverted microscope optical trapping setup using an 820 nm laser diode. The transmission of the laser diode through the double nanohole aperture is used to detect and monitor the trapping events since dielectric loading creates a large variation (typically around 10%) in the transmitted intensity. It should be noted that this method produces negligible heating due to the presence of a gold film; the heating is expected to be of the order of 0.1K.[11,12] Furthermore, the technique produces copious signal for only 3 mW of laser power, such that an optical density filter is used to avoid saturation of the avalanche photodiode.

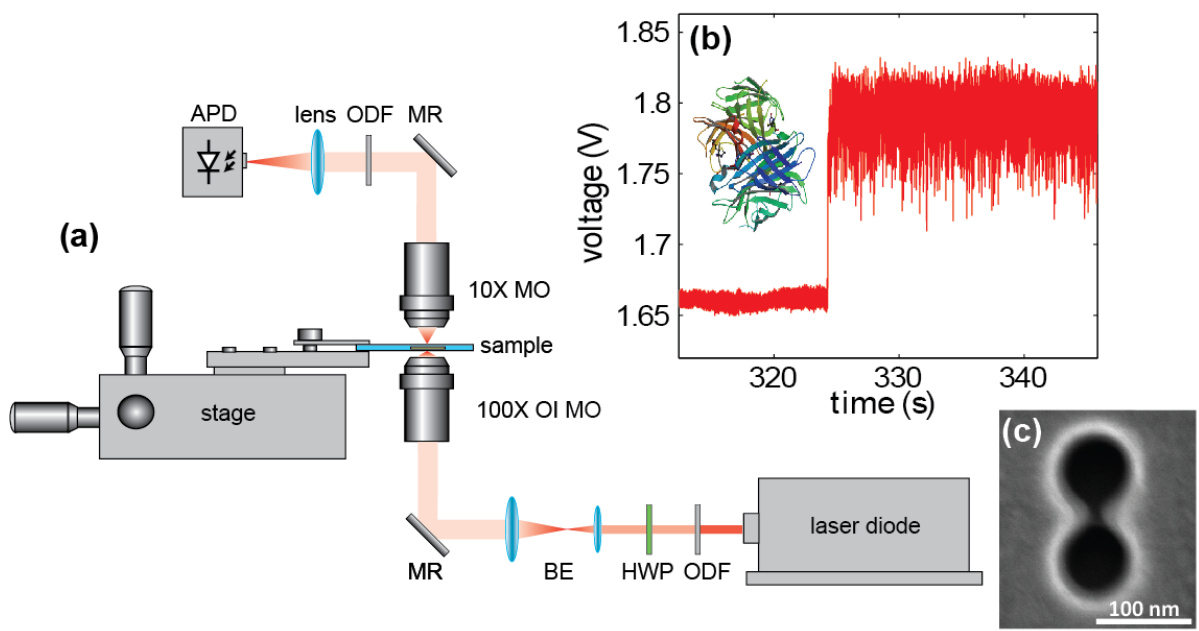

**Figure 1.** (a) A schematic of the double nanohole optical trap. Abbreviations used: ODF = optical density filter; HWP = half-wave plate; BE = beam expander; MR = mirror; MO = microscope objective; OI MO = oil immersion objective; APD = avalanche photodetector. (b) Optical trapping of biotinylated streptavidin (inset) seen as a sudden discrete jump in APD signal. (c) An SEM image of the double nanohole.

We prepared streptavidin solutions (Sigma Aldrich, 85878) in buffer with 0.01% w/v concentration. A portion of the solution was separated and exposed to excess biotin (Sigma Aldrich, B4501), which fully saturates the binding sites due to its high affinity. Fig. 2 shows the trapping dynamics of the solution with and without the biotin. The trapping event is seen by a discrete jump in the transmitted laser intensity, as denoted by arrows in Figs. 2(a) and 2(d). The streptavidin without biotin shows fluctuations in the transmitted intensity of the trapping laser with a timescale of about 600 ms, as seen in Figs. 2(b) and (c) (taken from different samples on different days). These fluctuations are absent in the biotinylated streptavidin, as seen in Figs. 2(e) and (f). An autocorrelation of the trapping events found for streptavidin and biotinylated streptavidin is shown in Fig. 3 which shows clearly the slower timescale dynamics of the streptavidin as compared to biotinylated streptavidin molecule.

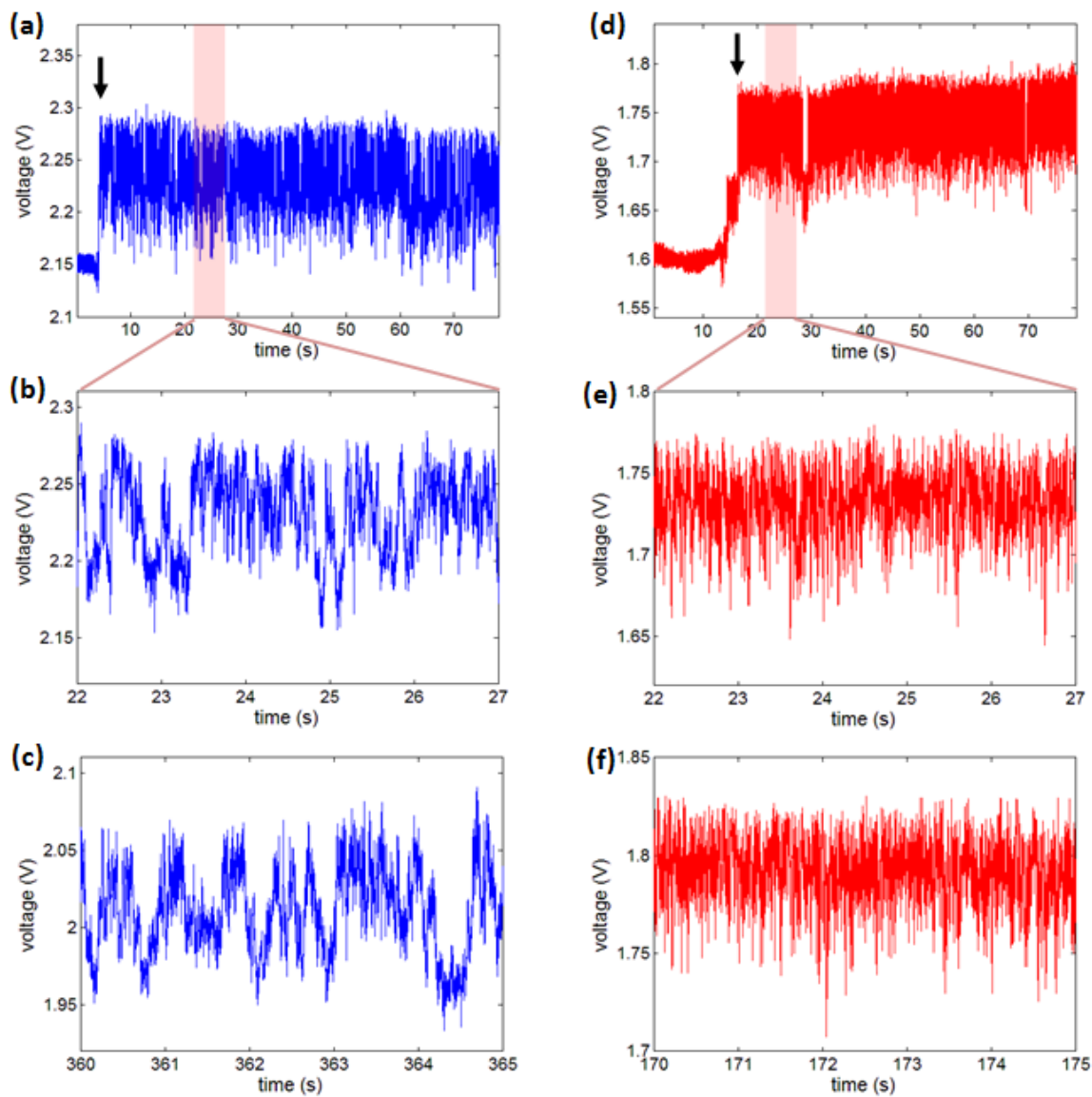

**Figure 2.** Trapping dynamics of streptavidin without and with biotin as measured from the APD voltage. (a) A time trace of a trapping event of a bare streptavidin molecule seen as an abrupt jump in the voltage level as denoted by the arrow. (b) Zoom-in of (a). (c) Repeat of (a) taken from a different sample on different day. (d) A time trace of a trapping event of a biotinylated streptavidin molecule seen as a discrete jump in the voltage level as indicated by the arrow. (e) Zoom-in of (d). (f) Repeat of (e) taken from a different sample on different day.

The change in the light transmitted through the aperture can arise from differences in the size of the particle trapped,[10] but also from changes in the particle shape or orientation, for example due to unfolding of proteins[9] or conformal changes. These marked differences are not expected to arise from mass-loading since biotin has a mass <0.5% the mass of streptavidin. The streptavidin without biotin shows fluctuations that are not present when biotin is added. This is consistent with numerical studies of streptavidin that suggest that the binding loop is highly mobile in the absence of biotin[13,14] – it would be interesting to attempt quantitative comparisons with molecular dynamics simulations in the future. The change in the optical transmission through the double nanohole aperture is expected to vary considerably due to stretching of the protein since the dipole moment increases along the axis of elongation. It is energetically favorable of the optical trap to stretch out the protein and this leads to more light transmitted through the double nanohole. Similar stretching behavior has been reported in other optical trapping systems, albeit for much larger particles such as cells.[15] Figure 2(d) shows occasional steps observed in the biotinylated-streptavidin (for example, at time 29 s), which suggests that even with small molecule binding, the protein can change its shape or orientation. Further investigation, perhaps comparing with numerical simulations, should help to clarify which process is taking place here.

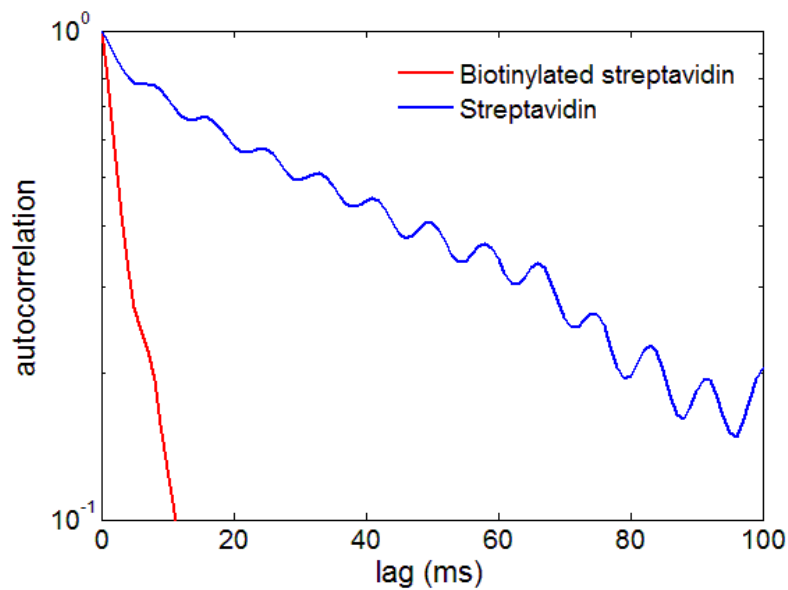

**Figure 3.** Autocorrelation Autocorrelation of trapped streptavidin APD signal fluctuations with and without biotin.

Future measurements are anticipated to observe the individual biotin binding events. Two approaches are considered. First, a flow-channel setup, similar to our past work using nanospheres,[16] may be used to first trap the vacant streptavidin and then introduce biotin into the channel. The goal here will be to determine if the binding to the four binding sites in streptavidin can be observed individually. Second, it is possible to repeat the experiments of this work with monovalent streptavidin to see if distinct dynamics are observed when only a single binding site is active.[17] Obviously, it is of great interest to see if these results can be extended to other PSMI systems that play a role in biological function. Also of interest is to see how scalable this approach is to enable multiplexed screening, for example by using multiple optical traps.[18,19]

In summary, we have shown that by studying the optical trapping dynamics on a single protein we can easily distinguish between the biotinylated and bare forms of streptavidin. Our approach does not require surface immobilization or exogenous markers, and it gives the real-time dynamics of individual protein molecules, representing the ultimate practical limit for sensitivity. While our work only uses the model biotin-streptavidin system, this work shows great potential for applications to screening small molecule drug candidates by monitoring their influence on proteins of interest,[1] and for the understanding the mechanisms of PSMIs.[20]

This work is funded by the NSERC Discovery Grant. The authors declare that there is no conflict of interest.

AUTHOR INFORMATION

**Corresponding Author**

*E-mail: rgordon@uvic.ca